\documentclass[english,twocolumn,pra,aps,showpacs]{revtex4}

\usepackage[dvips]{graphicx}
\usepackage{babel,amsmath,amssymb,float}
\usepackage{psfrag}
\usepackage{pstricks}
\usepackage{pst-node}
\usepackage{pst-plot}

\newcommand{\qed}{\hspace*{\fill}$\square$}

\newcommand{\be}{\begin{equation}}
\newcommand{\ee}{\end{equation}}

 \newcommand{\R}{\mathbf{R}}
 \newcommand{\C}{\mathbf{C}}
 
 \newcommand{\Z}{\mathbf{Z}}

 \newcommand{\e}{{\mathrm e}}

 \newcommand{\paratodo}{\forall\,}
 
 \newcommand{\vect}[1]{\boldsymbol{\mathrm{#1}}}
 
 \newcommand{\sset}[1]{ \{#1\} }

 \newcommand{\prima}{^\prime}

 \newcommand{\ket}[1]{|#1\rangle}
 \newcommand{\bra}[1]{\langle #1|}
 \newcommand{\braket}[2]{\langle #1|#2\rangle}

 \newcommand{\cluster}{\ket{\kappa}}
 
 \newcommand{\tcc}{\Psi_{\rm c}}

\newcommand{\fe}{\mathfrak e} %
\newcommand{\fv}{\mathfrak v} %
\newcommand{\ff}{\mathfrak f} %
\newcommand{\fu}{\mathfrak{u}} %
\newcommand{\fG}{\mathfrak G} %
\newcommand{\fU}{\mathfrak{U}} %
\newcommand{\fC}{\mathfrak{C}} %
\newcommand{\fV}{\mathfrak{V}} %
\newcommand{\fE}{\mathfrak{E}} %
\newcommand{\fF}{\mathfrak{F}} %
\newcommand{\fM}{\mathfrak{M}} %

\newcommand{\id}{1\hspace{-.25em}{\rm I}} %
\newcommand{\stabilizer}{{\mathcal S}} %

\begin{document}

\title[Short Title]{Statistical Mechanical Models
and Topological Color Codes}

\author{H. Bombin and M.A. Martin-Delgado}
\affiliation{ Departamento de F\'{\i}sica Te\'orica I, Universidad
Complutense, 28040. Madrid, Spain. }

\begin{abstract}
We find that the overlapping of a topological quantum color code
state, representing a quantum memory, with a factorized state of
qubits can be written as the partition function of a 3-body
classical Ising model on triangular or Union Jack lattices. This
mapping allows us to test that different computational capabilities
of color codes correspond to qualitatively different universality
classes of their associated classical spin models. By generalizing
these statistical mechanical models for arbitrary inhomogeneous and
complex couplings, it is possible to study a measurement-based
quantum computation with a color code state and we find that their
classical simulatability remains an open problem. We complement the
meaurement-based computation with the construction of a cluster
state that yields the topological color code and this also gives the
possibility to represent statistical models with external magnetic
fields.
\end{abstract}


\maketitle

\section{Introduction}
\label{sect_intro}

Recently, a very fruitful relationship has been
established between partition functions of classical
spin models and a certain class of quantum stabilizer
states with topological protection \cite{ndb_07a}, \cite{br_07}.
The topological quantum code states considered so far in these
studies correspond to the toric code states introduced by Kitaev
\cite{kitaev97}, \cite{dennis_etal02}. The classical spin model that emerges when
a planar toric code is projected onto a product state of single-qubits
with very specific coefficients is the standard classical Ising model
in two dimensions with homogeneous real couplings and zero magnetic
field.

Single-qubit measurements also appear naturally in a
measurement-based computation (MQC) scheme \cite{rauss_briegel_01},
\cite{rauss_briegel_01b}.
Thus, these connections between classical spin models and
topological quantum states are also useful to test whether those
topological states are efficiently classically simulable  with MQC.
It has been shown that MQC with a planar Kitaev code state as input
can be efficiently simulated in a classical computer if at each step
of the computation, the sets of measured qubits form simply
connected subsets of the two-dimensional lattice \cite{br_07}. The
connection of classical spin models with measurement-based quantum
computation has been shown to be useful to prove the completeness of
the classical 2D Ising model with suitably tuned complex
nearest-neighbor couplings in order to represent the  partition
function of the classical Ising model on arbitrary lattices, with
inhomogeneous pairwise interactions and local magnetic fields
\cite{ndb_07b}.

Topological color codes (TCC) were introduced to implement the set
of quantum unitary gates of the whole Clifford group by means of a
topological a stabilizer code in a two dimensional lattice
\cite{topodistill}, and then generalized to three dimensional
lattices in order to achieve a universal set of topological quantum
gates \cite{tetraUQC}.
These 2D and 3D realizations of TCC are instances of general
D-dimensional realizations. We call those lattices related to these
codes as D-colexes (for color complexes), and they are D-dimensional
lattices with coordination number $D+1$ and certain colorability
properties. Moreover, this codes can also appear as the ground state
of suitable Hamiltonians, and the corresponding quantum systems are
brane-net condensates. \cite{topo3D}.

Given these nice properties exhibited by the topological color
codes, it is natural to ask what type of classical spin models can
be constructed out of them and see whether they belong or not to the
same universality class of the classical Ising model arising in the
Kitaev model. In this work we address this issue and find that the
the overlapping of a TCC state with a product state of single qubits
with appropriate coefficients is mapped onto the partition function of
the 3-body classical Ising model on the dual lattice of the original
lattice where the color code is defined. For concreteness, we
consider the triangular and the
`Union Jack' lattice for these classical many-body spin systems. This represents
a sharp difference with the result obtained with the topological
states in the Kitaev code. In fact, the universality classes of the
3-body classical Ising model in several lattices are quite different
from the corresponding universality class of the standard 2-body
Ising model.

Moreover, we also study the topological color code states in
a MQC scenario to test their classical simulability.  We find that the
current state of knowledge in statistical mechanical models with
3-body interactions, arbitrary inhomogeneous complex couplings and
lattice shapes is much less developed than the 2-body Ising model
which is relevant for the case of the toric code states.
Thus, we conclude that the classical simulability
of TCC states with MQC remains an open problem.

In a MQC, the usual initial many-particle entangled state is a cluster
state \cite{rauss_briegel_01}, \cite{rauss_briegel_01b} instead of a
topological code. Then, we also show how to construct a color code state from
a certain cluster state. Interestingly enough, this construction
turns out to be useful for the description of statistical mechanical
systems with 3-body interactions in the presence of an external
field.

This paper is organized as follows:
in Sect.\ref{sect_II} we give an introduction to the
topological color code states needed to present
in Sect.\ref{sect_III} the mapping onto
the classical 3-body Ising model in the triangular and Union Jack lattices.
In Sect.\ref{sect_IV} we study the measurement-based quantum computation with
topologocial color code states by generalizing the results of the
previous section.
In Sect.\ref{sect_V} we show how to prepare a topological color code from a
cluster state as those introduced in MQC. This is also useful for studying
partition functions of statistical mechanical models with 3-body interactions
and external magnetic fields.
Sect.\ref{sect_conclusions} is devoted to conclusions.

\section{Topological Color Codes}
\label{sect_II}

\subsection{Construction}

Let us start by recalling the notion of a Topological Color Code
(TCC) in order to see what type of classical spin models we obtain
from them with appropriate projections onto factorized quantum states and specific
lattices.

\begin{figure}
\includegraphics[width=6 cm]{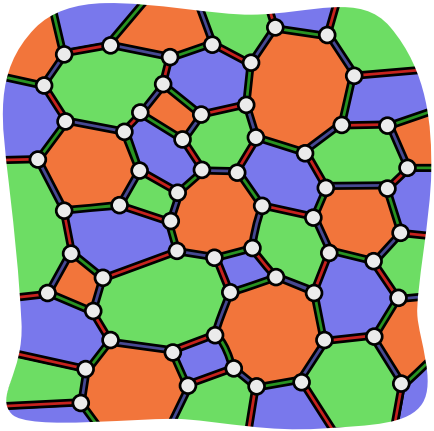}
\caption{\label{2_colex} An example of 2-colex. Both edges and face
are 3-colorable and they are colored in such a way that green edges
connect green faces and so on and so forth for red and blue edges
and faces.}
\end{figure}

A TCC, denoted by ${\cal C}$, is a quantum stabilizer error
correction code constructed with certain class of two-dimensional
lattices called 2-colexes. The word colex is a contraction that
stands for color complex, where complex is the mathematical
terminology for a rather general lattice \cite{homologicalerror}. A
2-colex, denoted by $\fC_2$, is a 2D trivalent lattice which has
3-colorable faces and is embedded in a compact surface of arbitrary
topology like a torus of genus $g$. A trivalent lattice is one for
which three edges meet at every vertex. The property of being
3-colorable means that the faces (or plaquettes) of the lattice can
be colored with these colors in such a way that neighboring faces
never have the same color. We select as colors red (r), green (g)
and blue (b). An example of a 2-colex construction is shown in
Fig.~\ref{2_colex}.

Edges can be colored in according to the coloring of the faces.
In particular, we attach red color to the edges that connect red
faces, and so on and so forth for the blue and gree edges/faces.
When studying higher dimensional colexes, it turns out that the
coloring of the edges is the key property of $D$-colexes: all the
information about a $D$-colex is encoded in its 1-skeleton, i.e.,
the set of edges with its coloring\cite{topo3D}.

\begin{figure}
\psfrag{ff}{$\ff$} %
\psfrag{g}{$\gamma$}%
\psfrag{1}{1} %
\psfrag{2}{2} %
\psfrag{3}{3} %
\psfrag{4}{4} %
\psfrag{5}{5} %
\psfrag{6}{6} %
\psfrag{a}{a} %
\psfrag{b}{b} %
\psfrag{c}{c} %
\psfrag{d}{d} %
\psfrag{e}{e} %
\psfrag{f}{f} %
\includegraphics[width=7cm]{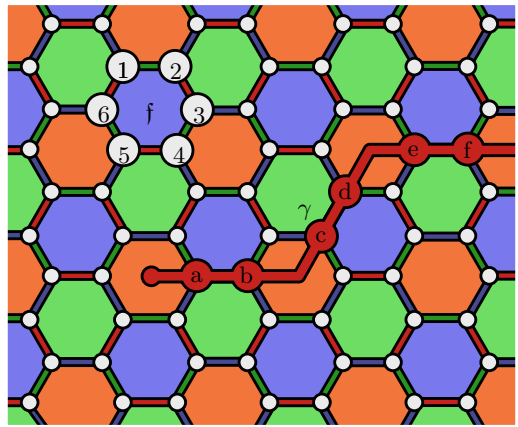}
\caption{\label{hexagonal} An hexagonal lattice is an instance of
2-colex. Numbered vertices belong to the face $\ff$. Vertices
labeled with letters correspond to the red string $\gamma$
displayed. $\gamma$ is an open string, because it has an endpoint in
a red face.}
\end{figure}

Given a 2-colex $\fC_2$, a TCC ${\cal C}$ is constructed by placing
one qubit at each vertice of the colored lattice. Let us denote
by $\fV$, $\fE$ and $\fF$ the sets of vertices $\fv$, edges $\fe$
and faces $\ff$, respectively, of the given 2-colex. Then, the
generators of the stabilizer group, denoted by $\stabilizer$, are
given by face operators only. For
each face $\ff$, they come into two types depending whether they are
constructed with Pauli operators of $X$- or $Z$-type: \be
\begin{split}
X_{\ff} &:=\bigotimes_{\fv \in \ff} X_{\fv}, \\
Z_{\ff} &:=\bigotimes_{\fv \in \ff} Z_{\fv},
\end{split}
\label{op_plaqueta} \ee and there are no generators associated to
lattice vertices. For example, an hexagonal lattice is an instance
of a 2-colex, see Fig.~\ref{hexagonal}. The operators for the face
$\ff$ displayed in the figure take the form $X_{\ff} =
X_1X_2X_3X_4X_5X_6, Z_{\ff} = Z_1Z_2Z_3Z_4Z_5Z_6$.
A given state $\ket{\Psi_{\rm c}}\in {\cal C}$ is left trivially
invariant under the action of the face operators, \be X_{\ff}
\ket{\Psi_{\rm c}} = \ket{\Psi_{\rm c}}, \ \ Z_{\ff} \ket{\Psi_{\rm
c}} = \ket{\Psi_{\rm c}}, \ \ \forall \ff \in \fF.
\label{code_states} \ee An erroneous state $\ket{\Psi}_{\rm e}$ is
one that violates conditions \eqref{code_states} for some set of
face operators of either type. As the generator operators
$X_{\ff}, Z_{\ff} \in {\stabilizer}$ satisfy that they square to
the identity operator, $(X_{\ff})^2 = \id =  (Z_{\ff})^2,
\forall \ff \in \fF$, then an erroneous state is detected by having
a negative eigenvalue with respect to some set of stabilizer
generators: $X_{\ff} \ket{\Psi}_{\rm e} = - \ket{\Psi}_{\rm e}$
and/or $Z_{\ff} \ket{\Psi}_{\rm e} = - \ket{\Psi}_{\rm e}$.

Interestingly enough, it is possible to construct a quantum lattice
Hamiltonian $H_{\rm c}$ such that its ground state is degenerate and
corresponds to the TCC ${\cal C}$, while the erroneous states are
given by the spectrum of excitations of the Hamiltonian
\cite{topodistill}. Such Hamiltonian is constructed out of the
generators of the topological stabilizer group $\stabilizer$, \be
H_{\rm c} =  - \sum_{\ff \in \fF} (X_{\ff} + Z_{\ff}).
\label{color_hamiltonian} \ee The ground state of this Hamiltonian
exhibits what is called a topological order \cite{wenbook04}, as
opposed to a more standard order based on an spontaneous symmetry
breaking mechanism. One of the signatures of that topological order
is precisely the topological origin of the ground state degeneracy:
the number of degenerate ground states depends on topological
invariants like Betti numbers \cite{topo3D}. In two dimensional
lattices, the relevant Betti number corresponds to the Euler
characteristic $\chi$ of the surface where the 2-colex is embedded.

\subsection{String-net operators}

In order to better understand both the ground state and excitations
of this Hamiltonian and their topological properties, it is rather
convenient to introduce the set of string operators that can be
defined on a 2-colex $\fC_2$. String operators are generalizations
of face operators \eqref{op_plaqueta} that can be either open or
closed, i.e., with or without end-points. These strings are
topological and like in Kitaev model, the homology is defined on
$\Z_2$ since we work with two-level quantum systems located at the
sites of the lattice. However, in a TCC we have an additional
ingredient to play around: color. Let us split the sets of edges and
faces into colored subsets denoted by $\fE := \fE_{r} \cup \fE_{g}
\cup \fE_{b}$ and $\fF := \fF_{r} \cup \fF_{g} \cup \fF_{b}$, where
$\fE_{r}$ is the subset of red edges, and similarly for the rest of
subsets.

A colored string $\gamma$ is a collection of edges of a given color.
Thus, a blue string $\gamma$ takes the form  $\gamma=\sset{e_i}$
with $e_i\in\fE_b$. The definition of colored string operators is
completely analogous to that of face operators: \be X_\gamma :=
\bigotimes_{\fe \in \gamma} X_{\fe},  \qquad Z_\gamma:=
\bigotimes_{\fe \in \gamma} Z_{\fe}, \label{color_strings} \ee
where, in turn, $ X_{\fe}=X_{\fv_1}\otimes X_{\fv_2}$ if $\fv_1$ and
$\fv_2$ are the sites at the ends of the edge $\fe$, and similarly
for $ Z_{\fe}=Z_{\fv_1}\otimes Z_{\fv_2}$. For instance, consider
the red string operator in Fig.~\ref{hexagonal}, where we have
$X_\gamma=X_aX_bX_cX_dX_eX_f\cdots,
Z_\gamma=Z_aZ_bZ_cZ_dZ_eZ_f\cdots$.
\begin{figure}
\includegraphics[width=7 cm]{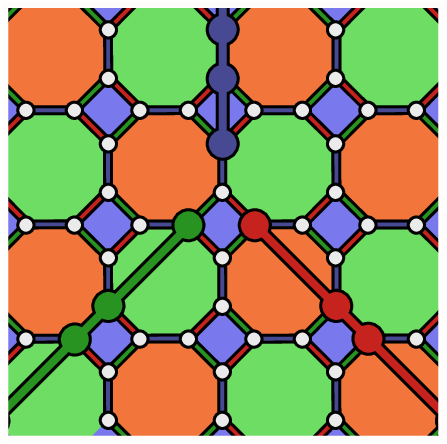}
\caption{\label{octogonal} A 4-8 lattice is an instance of 2-colex.
A closed string-net is displayed, composed of 3 strings of different
colors meeting at a branching point. The string-net is closed
because at each face we find an even number of its vertices.}
\end{figure}

Colored strings are open if they have endpoints. These endpoints are
localized at faces which share color with the string. In particular,
a face $\ff$ is an endpoint of $\gamma$ if the number of edges of
$\gamma$ meeting at $\ff$ is odd, see Fig.~\ref{hexagonal}. In terms
of string operators, a face $\ff$ is an endpoint of $\gamma$ if
$\sset{X_\gamma,Z_\ff}=0$ or, equivalently, if
$\sset{Z_\gamma,X_\ff}=0$. Thus, open string operators do not
commute with those face operators in their ends. In other words, a
string operator that commutes with all the face operators must
correspond to a closed string, that is, a string without endpoints.
In terms of the Hamiltonian \eqref{color_hamiltonian}, string
operators produce quasi-particle excitations at their ends when
applied to the ground state. These quasi-particle excitations are
Abelian anyons.

Closed strings are mainly of two types. They can be homologically
trivial, meaning that they are the boundary of certain area of the
surface, or homologically nontrivial. In terms of operators, a
string is a boundary iff its string operators belong to the
stabilizer group $\stabilizer$ of the color code $\cal C$. Such
boundary string operators are thus products of face operators. In
fact, face operators themselves are the basic boundary string
operators.

Although we have introduced colored strings and the corresponding
operators for illustrative purposes, in fact in a TCC we have to
deal with more general objects, namely string-nets. A string-net is
a collection of strings meeting at certain branching or ramification
points. An example of these types of configurations are shown in
Fig.~\ref{octogonal}.

A string-net $\gamma$ is a collection of vertices
$\gamma\subset\fV$. Equivalently, $\gamma$ is a formal sum of
lattice vertices $\fv \in \fV$ with coefficients $\gamma_{\fv}\in
\Z_2$, i.e., \be \gamma = \sum_{\fv \in \fV} \gamma_{\fv} \fv, \ee
where $\gamma_\fv=1$ if $\fv\in\gamma$ and $\gamma_\fv=0$
otherwise. Given a string-net $\gamma$, we define the string-net
operators \be X_{\gamma} := \bigotimes_{\fv} X^{\gamma_{\fv}},
\qquad Z_{\gamma} := \bigotimes_{\fv} Z^{\gamma_{\fv}}\ee Just as
in the case of colored strings, we can talk about open and closed
string-nets, and about trivial and non-trivial closed string-nets.
In terms of operators, the situation is exactly the same as with
strings. That is, a string-net $\gamma$ has an endpoint at a face
$\ff$ if $\sset{X_\gamma, Z_\ff}=0$, it is closed if its
string-net operators commute with all the face operators, and it
is a boundary if it is a product of face operators. In order to
translate these ideas into purely geometric terms, we can define the
boundary operator
\begin{equation}\label{boundary_color}
\partial_c \gamma := \sum_\ff x_\ff \ff, \quad x_\ff=
\begin{cases}
0, &|\gamma\cap\ff| \text{ is even,} \\
1, &|\gamma\cap\ff| \text{ is odd,}
\end{cases}
\end{equation}
where $|\gamma\cap\ff|$ is the number of vertices that $\gamma$ and
$\ff$ share. Thus $\partial_c\gamma$ is the formal sum of the
endpoints of $\gamma$. It is also natural to define an operator
$\partial_c$ for faces
\begin{equation}\label{boundary_color_faces}
\partial_c \ff := \sum_\fv x_\fv \fv, \quad x_\fv=
\begin{cases}
0, &\fv\not\in\ff, \\
1, &\fv\in\ff,
\end{cases}
\end{equation}
so that $\partial_c \ff$ is the string-net composed of the vertices
of $\ff$. With this definitions, $\gamma$ is closed if and only if
\begin{equation}\label{boundary_color_0}
\partial_c \gamma = 0,
\end{equation}
and it is a boundary if and only if there exist a collection of
faces $S=\sum_\ff S_\ff \ff$ such that
\begin{equation}\label{boundary_color_0b}
\gamma = \partial_c S.
\end{equation}

It is possible to give explicit expressions for the states of the
TCC or, equivalently, for the ground states of the Hamiltonian
\eqref{color_hamiltonian}. The states are superpositions of all
possible closed string-nets, a typical feature of the ground states
of systems with topological order \cite{wenbook04}, \cite{levinwen05}. The following is
an un-normalized ground state for any given 2-colex
\cite{topodistill}, \cite{topo3D} \be\label{color_gs}
\begin{split}
\ket {\Psi_{\rm c}} &:=  \prod_{\ff} (1+X_{\ff})
\,\, \ket 0^{\otimes |\fV|} \\
&= \sum_{\gamma\in \Gamma_0} X_{\gamma} \,\, \ket 0^{\otimes |\fV|}
=:  \sum_{\gamma\in \Gamma_0} \ket{\gamma},
\end{split}
\ee where $|\fV|$ is the number of vertices in the 2-colex $\fC_2$,
$\Gamma_0$ denotes the set of boundary string-nets and $\ket{0}$ is
the eigenstate $Z\ket{0}=\ket{0}$.

The degeneracy of the ground state or, equivalently, the number of
logical states encoded in the color code, depends on the topology of
the lattice. For a general 2-colex $\fC_2$ with Euler characteristic
$\chi(\fC_2):= |\fV| - |\fE| + |\fF|$, the number $k$ of encoded
qubits is given by $k=4-2\chi(\fC_2):= 2h_1$ \cite{topodistill},
where $h_1$ is the first Betti number of the surface where the
2-colex is embedded \cite{topo3D}. These additional ground states
can be obtained from the one given by \eqref{color_gs} by the action
of the encoded logical operators $\bar{X}_i, \bar{Z}_i$ with
$i=1,\ldots,k$. These, in turn, take the form of string-net
operators of non-trivial closed string-nets \cite{topodistill}.

For the purpose of this work, we shall be interested only in a
representative ground state like \eqref{color_gs}. Thus, we will
have to consider suitable surface topologies such that the
corresponding TCC is unique. We will return over this issue later
when we consider particular lattices.

\section{Connection with Classical Spin Systems}
\label{sect_III}

\subsection{Overlap and Partition Function}

Now, we come to the issue of what type of classical spin models may
arise from the color code state \eqref{color_gs} when we project it
onto a product state of a number of qubits given by $|\fV|$. In this
section we shall not consider the most general factorized state, but
one specifically adapted for the purpose of this connection in its
most simple form, namely,
\be \ket{\Phi_{\rm P}} := \bigotimes_{\fv \in \fV}
\ket{\phi}_{\fv}; \ \ \ket{\phi}_{\fv}:= \cosh (\beta J)
\ket{0}_{\fv} + \sinh (\beta J) \ket{1}_{\fv}, \label{factor_state1}
\ee
with $\beta:=1/k_{\rm B}T$ the inverse temperature parameter.

The classical spin model arises when computing the overlapping
between the ground state of the color code Hamiltonian
\eqref{color_gs} and this factorized state \eqref{factor_state1},
\be O(\beta J) := \langle\Psi_{\rm c}|\Phi_{\rm P}\rangle.
\label{overlapping} \ee Using \eqref{color_gs} and
\eqref{factor_state1} we get the following expression for this
overlapping,
\be
 O(\beta J) = \sum_{\gamma\in \Gamma_0} \bra{\gamma}
\bigotimes_{\fv \in \fV} \ket{\phi}_{\fv} = (\cosh (\beta J))^{|\fV|}
\sum_{\gamma\in \Gamma_0} u^{|\gamma|}, \label{overlapping_2}
\ee
$u:=\tanh (\beta J)$ and $|\gamma|$ is the number of vertices
of $\gamma$.

We want to relate \eqref{overlapping_2} to the partition function of
a classical spin system. So let $\fC_2$ be an arbitrary 2-colex.
Consider the dual lattice $\Lambda$. The vertices of $\Lambda$
correspond to the faces of $\fC_2$, and the faces of $\Lambda$ are
vertices in $\fC_2$. In particular, $\Lambda$ is a lattice in which
all faces are triangular and vertices are 3-colorable. Moreover, for
any such lattice $\Lambda$ there exist a suitable dual 2-colex
$\fC_2$.

So let us associate a classical system to $\Lambda$ by attaching
classical spin variables $\sigma_i=\pm 1$ to each of its sites $i$
(equivalently, to each face $\ff$ of $\fC_2$). The classical Hamiltonian is
\be {\cal H} := - J\sum_{\langle i,j,k\rangle} \sigma_i \sigma_j
\sigma_k,
\label{E_configuration}
\ee
where $J$ is a coupling
constant and the sum $\sum_{\langle i,j,k\rangle}$ is over all
triangles with spins $\sigma_i \sigma_j \sigma_k$ at their vertices.
Thus, we have a classical Ising model with 3-body interactions. The
case $J>0$ corresponds to a ferromagnetic model with an even parity
to be discussed below, and similarly $J<0$ to an antiferromagnetic
model with odd parity. The partition function of the model is
\be
{\cal Z} (\beta J) := \sum_{\{ \sigma \}} {\rm e}^{\beta
J\sum_{\langle i,j,k\rangle} \sigma_i \sigma_j \sigma_k},
\label{3body_t}
\ee
where the sum $\sum_{\{ \sigma \}}$ is over all
possible configurations of spins. The point then is that we have
\begin{equation}\label{igualdad_O_Z}
\mathcal Z (\beta J) = 2^N O(\beta J),
\end{equation}
where $N$ is the number of sites.

\begin{figure}
\psfrag{(a)}{(a)}%
\psfrag{(b)}{(b)}
\includegraphics[width=8 cm]{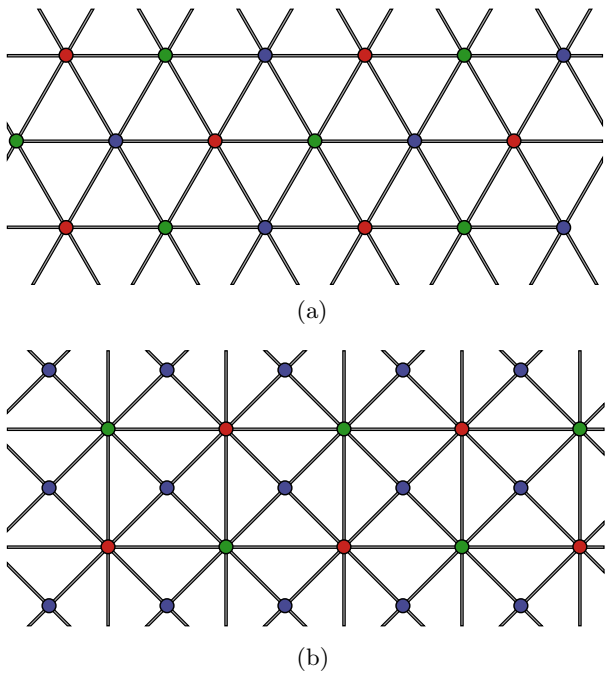}
\caption{Two instances of dual lattices of a 2-colex, which have
triangles as faces and have 3-colorable sites. The triangular
lattice (a) is dual to the hexagonal one. The Union Jack lattice (b)
is dual to the square-octogonal or 4-8 lattice.}
\label{figura_hex_UJ}
\end{figure}

Before we show why this identity holds, let us give a pair of
representative examples of dual lattices $\fC_2$ and $\Lambda$.
First, if the 2-colex is an hexagonal lattice then the dual lattice
$\Lambda$ is a \emph{triangular lattice},
Fig.~\ref{figura_hex_UJ}(a). Second, if the 2-colex is a
square-octogonal lattice (also denoted by 4-8 lattice), then its
dual is a \emph{Union Jack lattice}, see
Fig.~\ref{figura_hex_UJ}(b). The relevance of these examples is
two-fold. On the one hand, the hexagonal lattice is the simplest
lattice for a 2-colex and the 4-8 lattice is the simplest one when
we want to obtain TCC with certain transversality properties
for quantum computation (see
below). On the other hand, 3-body classical Ising-models on both
lattices have been studied in statistical mechanics to some extent.

\begin{figure}
\includegraphics[width=8 cm]{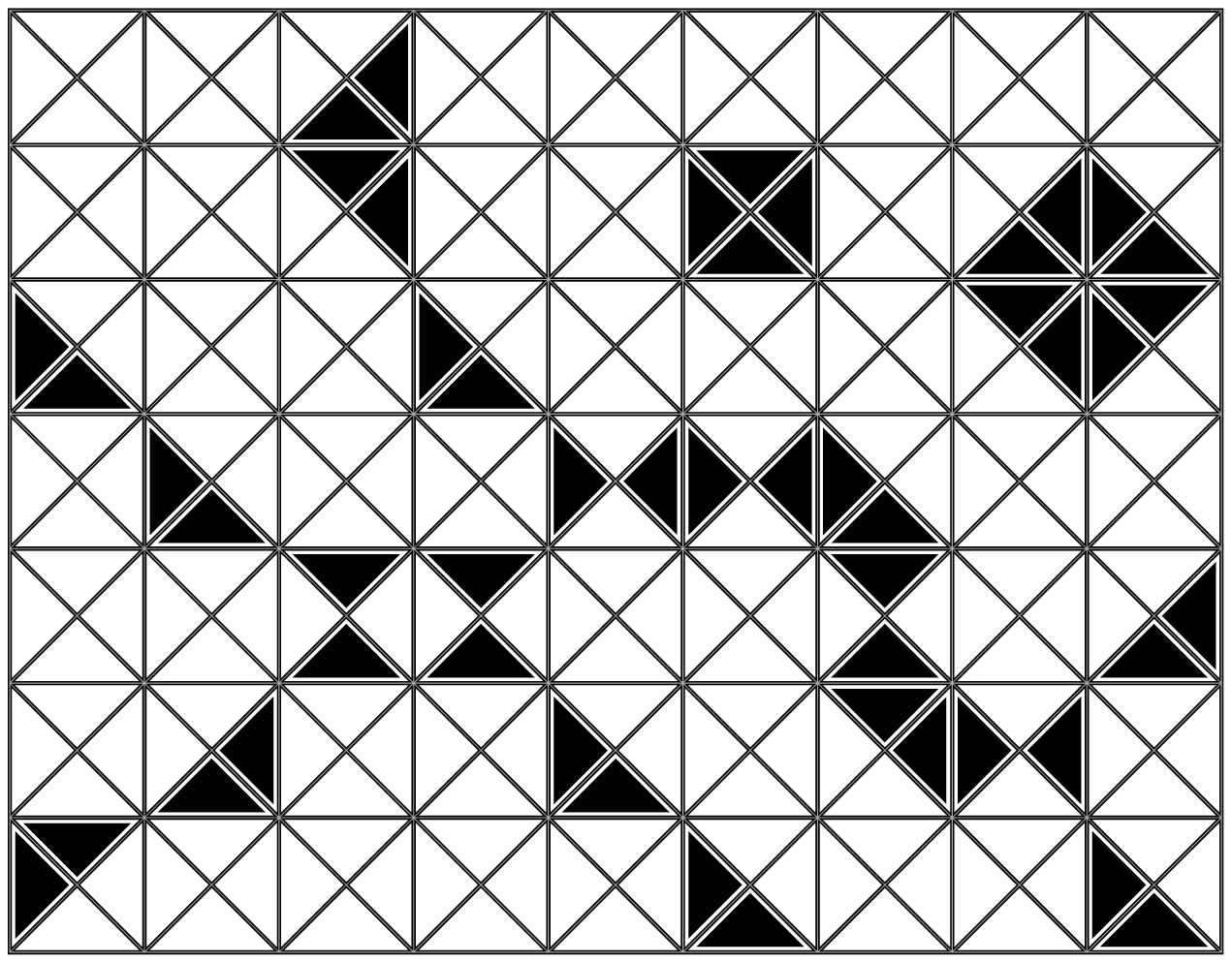}
\caption{A typical chain of triangles $\delta\in\Delta_0$ in a
triangular lattice. It is understood that only part of the lattice
is displayed. Black triangles represent the elements of $\delta$.
The fact that $\delta\in\Delta_0$ means that at each vertex always
meet an even number of triangles. }
\label{figura_triangulos}
\end{figure}

To prove \eqref{igualdad_O_Z}, let us start expanding the partition
function ${\cal Z} (\beta J)$ \eqref{3body_t} using the following
identity,
\be {\rm e}^{\beta J\sigma_i \sigma_j \sigma_k} = \cosh
(\beta J) + \sigma_i \sigma_j \sigma_k \sinh (\beta J).
\ee
Inserting it into \eqref{3body_t}, we may expand the partition
function as
\begin{equation}\label{desarrollo_particionS}
{\cal Z} (\beta J) = (\cosh (\beta J))^N \sum_{\{ \sigma\}}
\prod_{\langle i,j,k\rangle} (1 + u \sigma_i \sigma_j \sigma_k).
\end{equation}
Let us rewrite \eqref{desarrollo_particion} in the form
\begin{equation} \label{desarrollo_particionS2}
{\cal Z} (\beta J) = (\cosh (\beta J))^N \sum_{\delta}
u^{|\delta|} \sum_{\{ \sigma\}} \prod_{\langle i,j,k\rangle} (
\sigma_i \sigma_j \sigma_k)^{\delta_{ijk}},
\end{equation}
where $\delta = \sum_{\langle
i,j,k\rangle}\delta_{ijk}\triangle_{ijk}$ is a chain of triangles,
that is, a formal sum over triangles with binary coefficients, and
$|\delta|$ is the number of triangles in $\delta$. Using the
identities \be \sum_{ \sigma=\pm 1}\sigma^{n_o}=0, \ \ \sum_{
\sigma=\pm 1}\sigma^{n_e}=2, \label{binary_conditions} \ee where
$n_o$ and $n_e$ are odd and even numbers, respectively, we get
\begin{equation} \label{desarrollo_alta_T_S}
{\cal Z} (\beta J) = (2\cosh (\beta J))^N \sum_{\delta\in\Delta_0}
u^{|\delta|},
\end{equation}
where $\Delta_0$ contains those chains of triangles such that at any
given site $i$ an even number of triangles meet, as shown in
Fig.~\ref{figura_triangulos}. In fact, this type
of expansion is called a high-temperature expansion of the partition
function of a statistical mechanical model
\cite{newell_montroll_53}.

In order to compare \eqref{overlapping_con_campo} and
\eqref{desarrollo_alta_T}, we simply observe that triangles
$\triangle_{ijk}$ in $\Lambda$ correspond to vertices of the 2-colex
$\fv\in\fV$. This correspondence relates in an obvious way a
string-net $\gamma$ with a triangle chain $\delta$, in such a way
that $\Delta_0$ is identified with $\Gamma_0$. Therefore, we have
the desired relationship between the overlapping and the partition
function \eqref{igualdad_O_Z}.

 Although the previous
derivation was performed for a model with uniform couplings, it is
possible to obtain a completely analogous result for triangle
dependent couplings $J_{ijk}$. For simplicity we have preferred to
do the exposition with uniform couplings because, in fact, the case
of non-uniform couplings is contained in the more general case of
non-uniform couplings with non-uniform external field, to be
considered in section~\ref{sect_V}.

\subsection{Consequences}

We hereby draw a series of very important consequences from these
results, that will continue in the next section.

\noindent {\bf i/ Interactions:} We see that there is a clear
qualitative difference between topological color code states and
Kitaev's toric code states since they yield quite different type of
spin interactions:  a TCC state yields a 3-body interaction like in
\eqref{3body_t}, while a Kitaev's code produces the standard 2-body
classical Ising model, namely,
\be {\cal Z_{\rm Ising}} (\beta J) :=
\sum_{\{ \sigma \}} {\rm e}^{\beta J\sum_{\langle i,j\rangle}
\sigma_i \sigma_j}.
\label{2body}
\ee

\noindent {\bf ii/ Symmetry:} A distinctive feature of our result
\eqref{3body_t} is that the classical spin model associated to the
TCC state does not posses the up-down $\Z_2$ spin-reversal symmetry.
However, the partition function \eqref{3body_t} exhibits a
$\Z_2\times \Z_2$ symmetry. Recall that the lattice $\Lambda$ is
3-colorable at sites, so that we can redundantly label our classical
spin variables as $\sigma_i^c$ with $c=r,g,b$ the color at site $i$.
Then the change of variables
\begin{equation}
\sigma_i^c \rightarrow s(c) \ \sigma_i^c,\qquad s(r)s(g)s(b)=1,
s(c)=\pm 1,
\end{equation}
gives a global symmetry, which has symmetry group $\Z_2\times \Z_2$
because $s(b)=s(r)s(g)$. The ground states have to display this
symmetry, in fact. Consider states in which the values of the spin
variables only depend on the color, that is, for which
$\sigma_i^c=f_c$ with $f_c=\pm 1$. Such states can be labelled with
the tag $(f_r,f_g,f_b)$. Then it is easy to check that for the
ferromagnetic case $J>0$ the ground states are labelled with the
positive parity tags $(+,+,+), (+,-,-), (-,+,-), (-,-,+)$ whereas in
the antiferromagnetic case $J<0$ they are labelled with the negative
parity tags $(-,-,-), (-,+,+), (+,-,+), (+,+,-)$. Thus, each parity
sector, or classical ground state of \eqref{E_configuration} is
fourth-fold degenerate. Notice that the 3-body Ising model in such
3-colorable lattices of triangles shows no frustration, as opposed
to the standard Ising model \eqref{2body} in such lattices which is
indeed frustrated.

Remarkably, the gauge group underlying the topological order related
to Hamiltonian \eqref{color_hamiltonian} is also $\Z_2\times\Z_2$.
The situation is the same also with toric codes, where the global
symmetry of the classical system is $\Z_2$ and the gauge group for
the toric code topological order is $\Z_2$. This is certainly not a
matter of chance, since one can relate the types of domain walls in
the classical system to the types of condensed strings in the
quantum system.

\noindent {\bf iii/ Selfduality:} The models
in the triangular and Union Jack lattices
turn out to be self-dual like the usual 2-body Ising model, with a
critical temperature $\beta_c$ given by the same condition,
\be
\sinh 2K_c = 1, \ \ K_c:=\beta_c J_c = 0.4407.
\ee

Duality is a property between high-temperature and low-temperature
expansions of a statistical mechanical model like \eqref{3body_t} or
\eqref{2body}. A high-temperature expansion is a polynomial in the
variable $u=\tanh(\beta J)$ that is small when $T\rightarrow \infty$,
while a low temperature expansion is another polynomial in the
variable $u^{\ast}:={\rm e}^{-2\beta J^{\ast}}$ that is small in the
limit $T\rightarrow 0$. Then, a self-duality is a relationship between
the high-temperature expansion of one classical spin model in a
given lattice $\Lambda$ and the low-temperature expansion of the
same lattice. This is precisely the case of the 3-body Ising model
\eqref{3body_t} on both the triangular and Union Jack lattices
\cite{wood_g_72}, \cite{merlini_gruber_72} and the standard Ising
model \eqref{2body} \cite{krammers_wannier_41}.

\noindent {\bf iv/ Universality Classes:} Interestingly enough, the
3-body Ising model on the triangular lattice \cite{baxter_wu_73},
\cite{baxter_wu_74} and the Union-Jack lattice \cite{h_m_72} are
exactly solvable models under certain circumstances and this fact
allows us to check their criticality properties when compared with
those of the standard Ising model solution.

The critical exponent for the specific heat in the 3-body Ising model on the triangular
lattice is $\alpha=\frac{2}{3}$. This represents a power law behaviour which is in sharp contrast
with the well-known logarithmic divergence ($\alpha=0$) of the specific heat in the standard Ising model
\eqref{2body}. Other representative exponents are also different: the correlation length
exponent is $\nu=\frac{2}{3}$ (vs. $\nu=1)$, the magnetization exponent is $\beta=\frac{1}{12}$
(vs. $\beta=\frac{1}{8}$), while they share the same
two-point correlation function exponent at the critical
point $\eta=\frac{1}{4}$.

For the 3-body Ising model on the Union Jack lattice, the specific heat critical exponent is
also remarkably different $\alpha=\frac{1}{2}$. In fact, if the coupling constant $J$ is allowed to be anisotropic,
then even the critical exponent $\alpha$ may take on
a set of continuous values in $(0,\frac{1}{2})$ depending on a parameter related to
the coupling constants \cite{h_m_72}.

The computational capabilities of a topological color code depends
on the 2-colex lattices where it is defined. For a TCC on a
square-octogonal lattice it is possible to implement the whole
Clifford group of unitary gates generated by the set of gates $\{ H,
K^{\frac{1}{2}}, \Lambda_2\}$, where $H$ is the Hadamard gate,
$K^{\frac{1}{2}}$ the $\pi/8$-gate and $\Lambda_2$ the CNOT gate
\cite{topodistill}. However, for a 2-colex like the hexagonal
lattice the set of available gates is more reduced since the
$\pi/8$-gate cannot be implemented topologically \cite{topodistill}.
Thus we point out a remarkable connection between different
computational capabilities of color codes that correspond to
qualitatively different unversality classes of their associated
classical spin models, despite the fact that both color codes have
the same topological order.

\subsection{Borders}

If we want to consider classical systems of spins with a finite
number of sites, then we have to introduce either borders or a
nontrivial topology. Since in TCCs the nontrivial topology gives
rise to degeneracy, it is preferable to have borders. Also, borders
play a role for the ideas to be explained in the next section.

\begin{figure}
\includegraphics[width=8 cm]{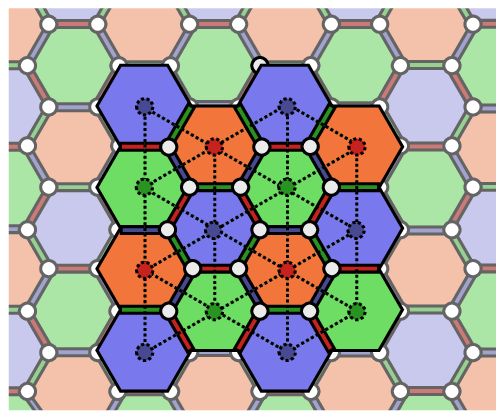}
\caption{Here both an hexagonal 2-colex $\fC_2$ and its dual
triangular lattice $\Lambda$ (dashed) are displayed to illustrate
how borders are introduced in the 2-colex if $\Lambda$ has borders.
All vertices of the 2-colex which are not triangles in $\Lambda$
have been removed, and also all the faces which keep no vertices.
The faces that only keep part of their vertices remain, but only
their $Z$ face operators are kept in the stabilizer.}
\label{figura_borde}
\end{figure}

In TCCs, borders can be of several types. For example, in
\cite{topodistill} it was shown how to build borders of a given
color. Here our guide to construct the border must be the dual
lattice $\Lambda$, which now has a border, along with the properties
of the classical system. Then, as shown in Fig.~\ref{figura_borde},
in order to construct the stabilizer for a TCC with border in such a
way that $\eqref{igualdad_O_Z}$ remains true is to start with an
infinite 2-colex $\fC_2$ and then keep only part of it following
certain criteria. (i) Keep those vertices $\fv$ of $\fC_2$ which
correspond to triangles in $\Lambda$. (ii) For those faces $\ff$ of
$\fC_2$ which keep all their vertices, we keep the face operators
$X_\ff$ and $Z_\ff$. (iii) For those faces $\ff$ which only keep a
subset $\ff\prima$ of their vertices, we introduce a face operator
$Z_{\ff\prima}$ acting on those qubits. Condition (i) ensures the
correspondence between triangle chains in $\Lambda$ and string-nets
in $\fC_2$. Conditions (ii) and (iii) ensure the correspondence
between $\Gamma_0$ and $\Delta_0$.

\section{Measurement-Based Quantum Computation with Color Codes}
\label{sect_IV}

In a measurement-based quantum computer (MCQ) \cite{rauss_briegel_01},
information processing is carried out via a sequence of one-qubit
measurements on an initialized entangled  quantum register. This is
an alternative to the standard gate-based quantum computation that
can simulate quantum networks efficiently.

An interesting problem is to study the performance of the MQC when
the initial entangled multiparticle state is a topological toric
code like in Kitaev's model \cite{br_07}. In particular, under which
general circumstances the MQC based on the planar Kitaev code can be efficiently
simulated by a classical computer. The answer to this question is that
the planar code state can be efficiently simulated on a classical computer
if at each step of MQC,  the sets of measured and unmeasured qubits
correspond to simply connected subsets of the lattice \cite{br_07}.

Likewise, another very interesting problem is whether a topological color code
state is classically sumulable in an scenario of measurement-based
quantum computation (MQC). By extending the results of sect.~\ref{sect_III},
it is possible to address this problem here.
Our aim is to see what conclusions can we learn from the statistical
mechanical models in order to test the classical simulability by MQC of the topological
code states.

The results of sect.~\ref{sect_III} can be interpreted as a complete projective
measurement of the planar topological color code state \eqref{color_gs}
onto a very specific product of single qubit states \eqref{factor_state1}.
This type of global measurements are not enough for doing a MQC based on the
color code state. Instead, we need to allow for more general one-qubit measurements
and to perform them in an adaptive fashion as the computation proceeds from the
starting point till the end.

To be more specific, let us consider a generic qubit state with complex coefficients
\be
\ket{\varphi}_{\fv}:= c^0_{\fv} \ket{0}_{\fv} + c^1_{\fv} \ket{1}_{\fv}, \ \
c^0_{\fv}, c^1_{\fv} \in \C.
\label{general_single_qubit}
\ee
Then, a MQC starts with a planar color code \eqref{color_gs} and we apply a series
of projective measurements $M_{\fv}:=\ket{x}_{\fv}\bra{x}, x=0,1$, from the first
qubit $\fv=1$ in the code until the last one $\fv=|\fV|$. The order in which this
sequence  of measurements is carried out through the 2-colex lattice is arbitrary.
After each measurement, the corresponding qubit at the vertex $\fv$ gets projected
onto one the states in \eqref{general_single_qubit} and the result is a value for
the outcome denoted by $m_{\fv}=0,1$.

The result of a run of a MQC is a set of outputs $m_1,\ldots, m_{\fv}, \ldots, m_{|\fV|}$
with a certain probability distribution $P(m_1,\ldots, m_{|\fV|})$.
We adopt the definition \cite{br_07} that a MQC is classically simulable
in an efficient way if there exists a classical randomized algorithm
that allows one to sample the outputs $m_1,\ldots, m_{|\fV|}$ from
the probability distribution $P(m_1,\ldots, m_{|\fV|})$ in a time poly$(|\fV|)$.

Then, after a complete run of a MQC with \eqref{color_gs} we get the following
generalized overlapping
\be
O_{\rm MQC} := \bra{\Psi_{\rm c}}\bigotimes_{\fv \in \fV}\ket{\varphi}_{\fv}.
\label{overlapping_MQC}
\ee
Using the same type of computations that led to \eqref{overlapping_2}, we arrive
at the following expression
\be
\begin{split}
O_{\rm MQC} &= \sum_{\gamma\in \Gamma_{0}} \bra{\gamma}
\bigotimes_{\fv \in \fV} \ket{\varphi}_{\fv} \\
&= \prod_{\fv \in \fV}c^{0}_{\fv} \sum_{\gamma\in \Gamma_{0}}
\prod_{\fv \in \fV: x_{\fv}=1} \left(\frac{c^{1}_{\fv}}{c^{0}_{\fv}}\right).
\end{split}
\label{overlapping_MQC2}
\ee
This result can also be turned into the partition function of a statistical model
with 3-body Ising interactions \eqref{3body_t} but with complex and inhomogeneous
Boltzmann weights
\be
w_{ijk} = {\rm e}^{\beta J_{ijk}} \in \C,
\label{complex_weights}
\ee
depending on each triangular plaquette $<i,j,k>$. The generalized overlapping
\eqref{overlapping_MQC2} is
proportional to a generalized partition function of a 3-body Ising model with
inhomogeneous and complex coupling constants
\be
{\cal Z} (\beta,\{ J_{ijk} \}) := \sum_{\{ \sigma \}}
{\rm e}^{\beta \sum_{\langle i,j,k\rangle\in \Lambda} J_{ijk} \sigma_i \sigma_j \sigma_k},
\label{3body_t_complex}
\ee
where the lattice $\Lambda$ can be the complete triangular or the complete
Union Jack lattice.

At an intermediate stage of a run of a MQC with the color code state, the 2-colex
will split into two subsets $\fC_2:= \fM \cup \overline{\fM} $, with $\fM$ the set of
measured qubits and $\overline{\fM}$ the set of unmeasured qubits \cite{br_07}. Thus, during the
running of the MQC starting with the color code state, we shall find generalized
partition functions of the type \eqref{3body_t_complex} but for a lattice that is
the dual of the subset of measured qubits: $\Lambda = \fM^{\ast}$.

Therefore, we arrive at the situation that in order to classically simulate a
topological color code state in a MQC we need to simulate the conditional probabilities
$P(m_{\fv+1}|m_1,\ldots, m_{|\fv|})$ (at step $\fv +1$ knowing the probabilities of previous steps)
and for these we need to be able to compute efficiently in a time poly(L)
on the size $L$ of the intermediate lattice at step $\fv +1$.

At this point there is a sharp difference between the classical simulation with MQC
of Kitaev states and color code states. Kitaev states can be classically simulated
under very general conditions: the subsets $\fM$ and $\overline{\fM}$ must be
simply connected.
The basic ingredient to achieve this result in the 2-body Ising model
is that even though a generalized standard Ising model (with arbitrary complex and
inhomogeneous couplings) may not be translationally invariant,
nevertheless there always exist a technique allowing it to be mapped onto a dimer
covering problem (DCP)
which in turn can be solved efficiently through the Paffian method in polynomial time
\cite{br_07}, \cite{barahona_82}, \cite{kasteleyn_61}.

However, the dimer problem technique is applicable to 2-body interactions but not for
the 3-body interactions that arise in the generalized statistical mechanical models from
color codes. In the case of the 3-body Ising model with uniform and
real couplings $J_{ijk}=J\in \R$ in a triangular lattice,
it can be exactly solved by mapping it onto the generating
function of a suitable site-colouring problem (SCP) on a hexagonal lattice \cite{baxter_wu_73}, \cite{baxter_wu_74},
which can be solved by the Bethe ansatz.
In order for this site-coloring mapping to work, certain restrictive conditions on the
triangular lattice must be fulfilled. In particular, the partition function
\eqref{3body_t} has to be defined on a periodic triangular lattice with $L$ rows
in the horizontal direction and $N$ columns in the vertical direction. Let us denote it
as $Z_{LN}^{\rm (3)}$. Let us also denote by $Z_{MN}^{\rm SCP}$ the generating function
of a site-coloring problem on a hexagonal lattice with $M=\frac{2}{3}L$ and $N$ columns.
Then, the aforementioned mapping works in the limit $N\rightarrow \infty$ as
\be
Z_{LN}^{\rm (3)} = Z_{MN}^{\rm SCP}, \ \ N\gg 1.
\label{scp_mapping}
\ee
Furthermore, the SCP is solved by Bethe ansatz technique. This technique
also poses another fundamental problem in this situation since it is used
to compute the eigenvalues of the associated transfer matrix of the SCP and then
the issue about the completeness of that spectrum in terms of Bethe ansatz
eigenfunctions arises. This issue is always a difficult question and, strictly speaking,
it is a conjeture.
Quite on the contrary,
these difficulties are absent in the standard Ising model case since the DCP
is more versatile.

The situation becomes even more difficult if we consider the generalized partition
function \eqref{3body_t_complex} in the framework of an intermediate step in
the MQC. Then, it is not known how to solve it efficently with a mapping to a SCP
in an hexagonal lattice without restrictions.

This site-coloring mapping plays a similar role
than the dimer covering mapping in the standard 2-body Ising model. However, the known solutions
to this site-coloring problem demand more restrictive conditions on the type of lattices and they
are less powerful than the dimer mapping technique.

As for the topological color code on a 2-colex like the Union Jack lattice,
similar conclusions apply: in the the case of real couplings $J_{ijk}=J\in \R$,
it is exactly solvable since it can be mapped onto a 8-vertex model \cite{h_m_72},
which in turn has to be solved by the Bethe ansatz.

The fact that the 3-body classical Ising model is exactly solvable
in the triangular and Union Jack lattices for real and isotropic
couplings, is not enough so as to conclude that the corresponding
topological color code states are classically simulable in a MQC
scenario. Therefore, the classical simulability of topological color
codes with MQC remains an open problem.

\section{Cluster States and Models with an External Field}
\label{sect_V}

In a MQC scheme of quantum computation, the input state
is called a cluster state \cite{rauss_briegel_01} which
is a rather general entangled state associated to a great
variety of graph states, i.e., states constructed from
qubit states located at the vertices of a lattice specified
by the incidence matrix of a graph. A very important property
of these cluster states is that they can be created efficiently
in any system with a quantum Ising-type interaction
between two-state particles in the specified lattice configuration.
In Sect.\ref{sect_IV} we have assumed that the input state for the
MQC was a TCC state, without caring about how it could be prepared.
Here we show how such a topological color code state can be obtained
from a appropriate cluster state. As a by-product, this construction
will turn out to be useful for obtaining classical Ising models in
an external magnetic field being associated to color code states.

\subsection{Cluster state formulation of TCC}

\begin{figure}
\includegraphics[width=8 cm]{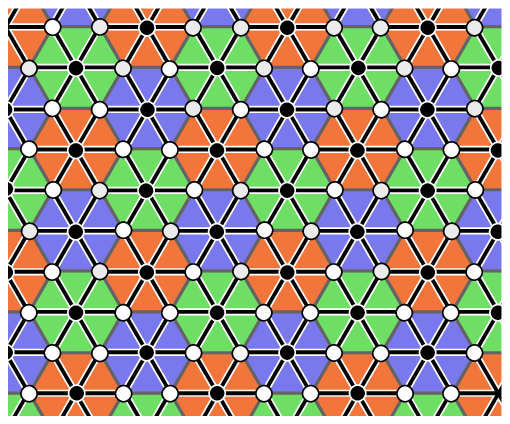}
\caption{The graph needed to obtain a color code state from a
cluster state. The graph is bipartite, and the vertices are divided
in black and white. Black vertices correspond to the faces of the
2-colex. White vertices correspond to the vertices of the 2-colex.
Black thick lines represent the edges of the graph, and grey lines
correspond to the edges of the 2-colex.} \label{figura_cluster}
\end{figure}

Instead of giving a general definition of cluster states, we will
consider only bipartite cluster states. So let $\fG$ be a finite
bipartite graph, that is, a graph such that its set of vertices
$\fU$ is the disjoint union of two sets $\fU=\fU_1\cup\fU_2$ in such
a way that neighboring vertices never belong to the same $\fU_i$.
Consider the quantum system obtained by attaching a qubit to each of
the vertices $\fu$. For each such vertex, we denote by $N(\fu)$ the
set containing both $\fu$ and its neighbors. The cluster state
$\cluster$ for the graph $\fG$ is then completely characterized by
the conditions:
\begin{align}\label{definicion_cluster}
\paratodo \fu\in\fU_1,\qquad &X_{N(\fu)}\cluster =\cluster,\nonumber\\
\paratodo \fu\in\fU_2,\qquad &Z_{N(\fu)}\cluster =\cluster.
\end{align}

In order to relate the TCC $\mathcal C$ of a 2-colex $\fC_2$ to a
cluster state, we construct a bipartite graph $\fG$ by setting
$\fU_1=\fV$ and $\fU_2=\fF$. Then the edges are defined so that
$\fu_1=\fv\in\fU_1$ is a neighbor of $\fu_2=\ff\in\fU_2$ if $\fv$ is
a vertex of $\ff$ in $\fC_2$, see Fig.~\ref{figura_cluster}. Observe
that the corresponding cluster state $\cluster$ satisfies
\begin{equation}
\paratodo \ff\in\fF, \qquad X_{\ff}\cluster =\cluster,
\end{equation}
because $X_{\ff}=\bigotimes_{\fv\in\ff} X_{N(\fv)}$. In fact, if
$\gamma$ is a string-net with $\partial_c \gamma =0$, then
$X_{\gamma}\cluster =\cluster$. However, to keep things as simple as
possible, let us just consider 2-colexes in which all closed
string-nets are boundaries. If we measure in the $Z$ basis all the
qubits corresponding to vertices $\fu\in\fU_2$ we will obtain a
series of binary values $x_\fu=x_\ff$. The remaining qubits are then
in a state characterized by the conditions:
\begin{align}
\paratodo \ff\in\fF,\qquad &X_{\ff}\cluster =\ket{\tcc (\vect x)},\nonumber\\
\paratodo \ff\in\fF,\qquad &Z_{\ff}\cluster = x_\ff \ket{\tcc (\vect x)}.
\end{align}
Thus, if $x_\ff=0$ for all the faces $\ff$, the result is the TCC
state $\ket\tcc$ \eqref{color_gs}. In the other cases the result is
essentially a TCC, in particular the state is
\begin{equation}
\ket{\tcc (\vect x)} := \sum_{\gamma\in\Gamma_{\vect x}} \ket\gamma
\end{equation}
where $\vect x$ denotes the binary vector of the measurement results
and $\Gamma_{\vect x}$ is the set of string-nets $\gamma$ with
$\partial_c\gamma=\sum_\ff x_\ff \ff$.

In fact, the original cluster state can be written as follows:
\begin{equation}
\cluster = \sum_{\vect x} \left( \ket{\vect
x}\otimes\sum_{\gamma\in\Gamma_{\vect x}} \ket\gamma\right)
\end{equation}
where $\ket{\vect x}=\bigotimes_{\fu\in\fU_2}\ket{x_\fu}_\fu$ is an
element of the computational basis of the subsystem of qubits in
$\fU_2$. To check that this is indeed the correct expression for the
cluster state, it is enough to note that it satisfies
\eqref{definicion_cluster}.

\subsection{Models with an external field}

Thus far, we have only considered statistical mechanical models
with zero external magnetic field. Here we go beyond that situation,
considering the formulation of models with 3-body Ising interactions
and arbitrary magnetic fields from the projection of topological
color codes onto appropriate product states.

Let us define the product state
\begin{equation}
\ket{\Phi_P (\vect J, \vect h)}:= \bigotimes_{\fv\in\fU_1} {\ket
{\phi(J_\fv)}}_{\fv} \bigotimes_{\ff\in\fU_2} {\ket
{\phi(h_\ff)}}_{\ff},
\end{equation}
where $\vect J=(J_\fv) \in \R^{|\fV|}$, $\vect h=(h_\ff) \in
\R^{|\fF|}$ and
\begin{equation}
\ket {\phi(s)} := \cosh s \,\ket 0 + \sinh s \,\ket 1, \quad s\in\R.
\end{equation}
Consider the overlapping
\begin{equation}\label{overlapping_con_campo_def}
O(\beta, \vect J, \vect h) := \braket {\tcc}{\Phi_p(\beta\vect J,
\beta\vect h)}.
\end{equation}
Its value is
\begin{multline}\label{overlapping_con_campo}
\sum_{\vect x}\bra {\vect x }\bigotimes_{\ff\in\fU_2} {\ket
{\phi(h_\ff)}}_{\ff}\sum_{\gamma\in\Gamma_{\vect x}}
\bra{\gamma}\bigotimes_{\fv\in\fU_1} {\ket
{\phi(J_\fv)}}_{\fv} = \\
\prod_\ff \cosh (\beta h_\ff) \prod_\fv \cosh (\beta J_\fv)
\sum_{\vect x}\sum_{\gamma\in\Gamma_{\vect x}}\prod_{\ff}
{u_\ff}^{x_\ff} \prod_{\fv} {u_\fv}^{\gamma_\fv},
\end{multline}
where
\begin{equation}
u_\ff:=\tanh (\beta h_\ff),\qquad u_\fv:=\tanh (\beta J_\fv).
\end{equation}

 We want to relate
\eqref{overlapping_con_campo_def} to the partition function of a
classical spin system. As in section~\ref{sect_III}, we consider the lattice
$\Lambda$ dual to the 2-colex $\fC_2$ and we associate a classical
system to $\Lambda$ by attaching classical spin variables
$\sigma_i=\pm 1$ to each of its sites $i$. This time however we
include triangle dependent couplings $J_{ijk}$ in triangles and a
site dependent external field $h_i$. Thus, we want to derive the
high temperature expansion for the partition function
\begin{equation}
{\cal Z} (\beta,\vect J, \vect h) := \sum_{\{ \sigma \}}
\e^{-\beta \cal H(\vect J, \vect h)},
\end{equation}
where the classical Hamiltonian is
\begin{equation}\label{Hamiltoniano_con_campo}
{\cal H}(\vect J, \vect h) := - \sum_i h_i\sigma_i -\sum_{\langle i,j,k\rangle}
J_{ijk}\sigma_i \sigma_j \sigma_k.
\end{equation}

We start using the identities
\begin{align}
\e^{\beta h_i\sigma_i} &= \cosh (\beta h_i) + \sigma_i
\sinh (\beta h_i),\nonumber\\
\e^{\beta J_{ijk} \sigma_i \sigma_j \sigma_k} &= \cosh (\beta
J_{ijk}) + \sigma_i\sigma_j \sigma_k \sinh (\beta J_{ijk}),
\end{align}
so that,
\begin{equation}\label{desarrollo_particion}
{\cal Z} (\beta,\vect J, \vect h) = C \sum_{\{ \sigma\}} \prod_i (1+u_i\sigma_i)
\prod_{\langle i,j,k\rangle} (1 + u_{ijk} \sigma_i \sigma_j
\sigma_k),
\end{equation}
where
\begin{equation}
C:=\prod_i \cosh (\beta h_i) \prod_{\langle i,j,k\rangle} \cosh
(\beta J_{ijk}),
\end{equation}
\begin{equation}
u_i:=\tanh (\beta h_i),\qquad u_{ijk}:=\tanh (\beta J_{ijk}).
\end{equation}
Let us rewrite \eqref{desarrollo_particion} in the form
\begin{equation} \label{desarrollo_particion2}
{\cal Z} (\beta, \vect J, \vect h) = C \sum_{\vect x}\sum_{\delta}\sum_{\{ \sigma\}}
\prod_{i} (u_i\sigma_i)^{x_i} \prod_{\langle i,j,k\rangle} (u_{ijk}
\sigma_i \sigma_j \sigma_k)^{\delta_{ijk}},
\end{equation}
where $\vect x=(x_i)$ is a binary vector and $\delta = \sum_{\langle
i,j,k\rangle}\delta_{ijk}\triangle_{ijk}$ is a chain of triangles,
that is, a formal sum over triangles with binary coefficients.
Reordering the expression we get
\begin{equation} \label{desarrollo_particion3}
{\cal Z} (\beta, \vect J, \vect h) = C \sum_{\vect x}\sum_{\delta}\epsilon(\vect x,
\delta)\prod_{i} {u_i}^{x_i} \prod_{\langle i,j,k\rangle}
{u_{ijk}}^{\delta_{ijk}} ,
\end{equation}
where
\begin{equation} \label{epsilon_x_delta}
\epsilon(\vect x, \delta) := \sum_{\{ \sigma\}} \prod_{i}
{\sigma_i}^{x_i} \prod_{\langle i,j,k\rangle} (\sigma_i \sigma_j
\sigma_k)^{\delta_{ijk}},
\end{equation}
From \eqref{binary_conditions} it follows that
\begin{equation} \label{valor_epsilon}
\epsilon(\vect x, \delta) = \left\{
\begin{array}{ll} 2^N, & \text{if } \forall i\,\,\prod_{\langle j,k\rangle_i} =
x_i,\\ 0, & \text{in other case.}\end{array} \right.
\end{equation}
where $\langle j,k\rangle_i$ denotes the pairs $(j,k)$ which form a
triangle with $i$, and $N$ is the number of sites. Finally we can
give the desired high temperature expansion of the partition
function:
\begin{equation} \label{desarrollo_alta_T}
{\cal Z} (\beta, \vect J, \vect h) = 2^N C \sum_{\vect x}\sum_{\delta\in\Delta_{\vect
x}}\prod_{i} {u_i}^{x_i} \prod_{\langle i,j,k\rangle}
{u_{ijk}}^{\delta_{ijk}},
\end{equation}
where $\Delta_{\vect x}$ contains those chains of triangles such
that at any given site $i$ an even (odd) number of triangles meet if
$x_i=0$ ($x_i=1$).

In order to compare \eqref{overlapping_con_campo} and
\eqref{desarrollo_alta_T}, we first observe that sites $i$
correspond to faces of the 2-colex $\ff\in\fF=\fU_2$ and triangles
$\triangle_{ijk}$ correspond to vertices of the 2-colex
$\fv\in\fV=\fU_1$. This correspondence relates in an obvious way
$x_i$ with $x_\fv$, $h_i$ with $h_\fv$ and so on and so forth. Also,
there is an exact correspondence between chains of triangle $\delta$
and string-nets $\gamma$. In particular, this correspondence
identifies $\Delta_{\vect x}$ with $\Gamma_{\vect x}$, so that we
get the desired relationship between the overlapping and the
partition function
\begin{equation}
\mathcal Z (\beta, \vect J, \vect h) = 2^N O(\beta, \vect J, \vect h).
\end{equation}

\vspace{0.1cm}

\section{Conclusions}
\label{sect_conclusions}

We have shown that the classical spin models associated
to quantum topological color code states in the  two dimensional
lattices called 2-colexes are Ising models
with 3-body interactions. We have studied this mapping in
the triangular and the Union Jack lattices, which are the
duals of the two very represenative 2-colexes, namely, the
hexagonal and the square-octogonal lattices, respectively.
This is a genuine difference with respect to the case with toric
code states which yield the partition function of
the standard Ising model with two-body interactions.
Ising models with different n-body interactions have very different
properties in general.
Remarkbly, different computational capabilities of the topological
color codes depending on the chosen 2-colex correspond to different
universality classes of the associated classical 3-body Ising models.

The tools employed to relate classical spin models with topological
color code states can be extended to study the performance of
such topological states when they are considered as input in a
measurement-based quantum computation. Then, the classical 3-body models
that arise involve arbitrary complex couplings and lattice shapes.
The problem of evaluating their corresponding generalized partition
functions cannot be performed with the dimer covering technique that
is so successful in the case of the classical 2-body Ising model.
The similar technique in the 3-body case is a particular site-coloring
problem that only in very specific instances has been solved by means
of the Bethe ansatz. The completeness of the Bethe ansatz poses in turn
more fundamental problems in this regard. Therefore, the fact that
the 3-body Ising model is exactly solvable in certain conditions is
not enough to conclude so far that the parent quantum topological
color color states are efficiently classically simulated by MQC.

Another interesting result is the construction of a cluster state
from which we can construct the topological color code state.
This turns out to be useful in order to obtain classical spin models
in the presence of arbitrary external magnetic fields.
Likewise, there are other two-dimensional multipartite states that
arise in the study of quantum antiferromagnets that may lead to
a variety of classical spin models \cite{rico_briegel_07}.

\noindent {\em Acknowledgements} We
acknowledge financial support from a PFI grant of the EJ-GV
(H.B.), DGS grants  under contracts BFM 2003-05316-C02-01,
FIS2006-04885 (H.B., M.A.M.D,),
and the ESF Science Programme INSTANS 2005-2010 (M.A.M.D.).


\end{document}